# On a possible melting curve of $C_{60}$ fullerite

**V. I. Zubov**[*,1], **C. G. Rodrigues**[2], and **I. V. Zubov**[1]

[1] Instituto de Física, Universidade Federal de Goiás, C.P. 131, 74001 Goiânia – GO,
Brazil and Department of Theoretical Physics, People's Friendship University, Moscow, Russia
[2] Núcleo de Pesquisa em Física da Universidade Católica de Goiás, Goiânia – GO, Brazil
`cloves@pucgoias.edu.br`

We study the thermodynamic properties of the high-temperature modification of fullerites on the basis of the Girifalco intermolecular potential. In the present work, using Lindemann's melting criterion, we estimate a possible melting curve $T_m(P)$ of $C_{60}$ fullerite. To take into account the lattice anharmonicity, which has a strong effect at $T > 700$ K, we use the correlative method of unsymmetrized self-consistent field. To check this approach, first we have applied it to solid Ar. In the range between its triple point $T_t = 83.807$ K and 260 K we obtained the mean square relative deviation from experimental data of about 0.7%. The melting curve for $C_{60}$ fullerite has been calculated from the melting point at normal pressure estimated at 1500 K up to 15 kbar, which corresponds to $T_m = 4000$ K, i.e. to the temperature estimated by Kim and Tománek [Phys. Rev. Lett. **72**, 2418 (1994)] as that of the decomposition of the $C_{60}$ molecule itself. The temperature dependence of the melting pressure is approximated very well by the Simon equation $\left(P_m(T)/\text{bar} - 1\right)/b = \left(T/T_0\right)^c$ with $T_0 = 1500$ K, $b = 6643.8$, and $c = 1.209$. The temperature dependence of the molar volume along the melting curve is described by $V_s(T) = V_s(T_0) - 29.20 \ln\left(T/T_0\right)$.

Мы изучаем термодинамические свойства высокотемпературной модификации фуллеритов на основе межмолекулярного потенциала Жирифалко. В настоящей работе, используя критерий плавления Линдеманна, оценивается возможная линия плавления фуллерита $T_m(P)$ $C_{60}$. Для учета ангармонизма колебаний решетки, который является сильным при $T > 700$ K, используется корреляционный метод несимметризованного самосогласованного поля. Для проверки используемого подхода он сначала был применен к твердому Ar. В интервале температур от тройной точки $T_t = 83.807$ K до 260 K получено среднеквадратичное отклонение от экспериментальных данных 0.7%. Линия плавления фуллерита $C_{60}$ рассчитана от нормальной точки плавления при нормальном давлении, оцененной в 1500 K до 15 кбар, что соответствует 4000 K, т.е. оцененной Кимом и Томанеком [Phys. Rev. Lett. **72**, 2418 (1994)] температуры распада самой молекулы. Температурная зависимость давления на линии плавления очень хорошо аппроксимируется уравнением Симона $\left(P_m(T)/\text{bar} - 1\right)/b = \left(T/T_0\right)^c$ с $T_0 = 1500$ K, $b = 6643.8$, $c = 1.209$. Температурная зависимость молярного объема вдоль этой линии описывается формулой $V_s(T) = V_s(T_0) - 29.20 \ln\left(T/T_0\right)$.

## 1 Introduction

Since the discovery of the fullerenes [1] and especially after the elaboration of effective methods for their preparation, separation and purification [2], which gave rise to their production in sufficient quantities for the growth of macroscopic crystals named the fullerites, they have been attracting a great deal of attention from scientists [3–6]. Among other things, the thermodynamic properties of fullerites have been investigated. Most of them are due to the lattice vibrations, whereas the dominant contribution to their specific heats comes from the intramolecular degrees of freedom.

After the preparation of fullerite crystals, a start has been made on the exploration of the phase transitions in these materials. The fullerene of the greatest abundance, $C_{60}$, and the next, $C_{70}$, have been studied most intensively. The phase transitions between the low-temperature, orientationally ordered, and high-temperature, disordered, modifications have been investigated both experimentally, e.g. [7–12], and theoretically [13]. Measurements of the saturated vapor pressures and the enthalpies of sublimation of these fullerites and also those of $C_{76}$ and $C_{84}$ have been made, and Markov et al. [14] have published summary data. In [15–17] one can find theoretical results.

The liquid phase of fullerites has never been observed. Nevertheless, discussion about its possible existence has persisted for years [18–27], based on a cluster-expansion-type method [18], an integral-equation approach [19], a Monte Carlo technique [20, 25, 26], a density-functional theory [21, 23], molecular-dynamics simulations [19, 24], a modified hypernetted-chain method [22] and also on the scaling of Lennard–Jones values [15, 19]. Hagen et al. [20] have reasoned that the liquid phase of $C_{60}$ has no region of absolute stability and hence cannot exist. However, other authors have shown the possibility of the liquid phase, although in a narrow phase diagram range. The estimations of its melting temperature (triple point) vary from 1400 to 1800 K. In our opinion, the upper values are closer to the spinodal point of the solid phase rather than to the melting temperature. Hasegawa and Ohno [25, 26] have also evaluated the liquid–vapor binodal curve including the critical point. They and Ferreira et al. [27] have also studied the solid–liquid coexistence but in a very narrow temperature interval. Note that Stetzer et al. [28] have reported that $C_{60}$ crystals heated at 1260 K for more than 10 min decomposed into amorphous carbon. However, this result has not been reproduced in other institutions, whereas the molecular-dynamics estimation for the decomposition temperature of a single $C_{60}$ molecule yields about 4000 K [29]. Because of this, investigations of the possibility of the liquid phase for fullerites are hitherto being continued [24, 27–29]. Recently Abramo et al. [30] have computed the phase diagrams for higher fullerites $C_{70}$, $C_{76}$, $C_{84}$ and $C_{96}$.

The present work is concerned with the evaluation of the pressure dependence of the possible melting temperature of $C_{60}$ fullerite up to 4000 K.

## 2 Calculation procedure

Owing to the lack of a unified rigorous microscopic theory for crystals and liquids, semi-empirical criteria for melting are of frequent use, which are stated as the constancy of one or other characteristic of a phase on the melting curve (a peculiar kind of 'integral of movement along the melting curve'). The first to be found was Lindemann's criterion, see e.g. [31]. It implies that on the melting curve

$$\delta = \sqrt{\overline{\boldsymbol{q}^2}}\big/a = \text{const}, \qquad (1)$$

where $\boldsymbol{q}^2 = 3\overline{q_\alpha^2}$ is the mean-square displacement of a molecule from its lattice points and $a$ is the nearest-neighbor distance. Although more recently other criteria have been established (the Ross criterion [32], the entropy [33] and energy [34] rules), interest has been shown in Lindemann's criterion up to present time [35].

Here we have applied it to the possible melting curve of $C_{60}$ fullerite. At $T > 700$ K, lattice anharmonicity has a strong effect on its properties. To take it into account, we use the correlative method of unsymmetrized self-consistent field (CUSF) [36]. Including anharmonic terms up to the fourth order in the zeroth approximation, the equation of state of the crystal at temperature $\Theta = kT$ under the pressure $P$ is of the form

$$P = -\frac{a}{3v}\left[\frac{1}{2}\frac{dK_0}{da} + \frac{\beta\Theta}{2K_2}\frac{dK_2}{da} + \frac{(3-\beta)\Theta}{4K_4}\frac{dK_4}{da}\right] + P^2 + P^H. \qquad (2)$$

Here $v(a)$ is the volume of the unit cell,

$$K_{2l} = \frac{1}{2l+1}\sum_{k\geq 1} Z_k \nabla^{2l}\Phi(R_k), \quad l = 0, 1, 2, \qquad (3)$$

$\Phi(r)$ is the intermolecular potential, $Z_k$ are the coordination numbers, $R_k$ are the radii of the coordination spheres, $\beta\left(K_2\sqrt{3/\Theta K_4}\right)$ is the solution of the transcendental equation

$$\beta = 3X\frac{D_{-2.5}(X+5\beta/6X)}{D_{-1.5}(X+5\beta/6X)}, \qquad (4)$$

in which $D_\nu$ are the parabolic cylinder functions, and $P^2$ and $P^H$ are corrections of the perturbation theory that make the effect of the anharmonicity more accurate.

For the intermolecular forces we use the Girifalco potential [37]

$$\Phi_G(r) = -\alpha\left(\frac{1}{s(s-1)^3}+\frac{1}{s(s+1)^3}-\frac{2}{s^4}\right) + \beta\left(\frac{1}{s(s-1)^9}+\frac{1}{s(s+1)^9}-\frac{2}{s^{10}}\right) \qquad (5)$$

where $s = r/2a$, $a = 3.55\times 10^{-8}$ cm is the effective hard-core radius of the molecule, and the coefficients $\alpha$ and $\beta$ are: $\alpha = 7.494\times 10^{-14}$ erg, and $\beta = 1.3595\times 10^{-16}$ erg. It has the minimum point $r_0 = 10.0558$ Å and the depth of the potential well is $\varepsilon/k = 3218.4$ K.

The mean-square molecular displacement in strongly anharmonic crystals is expressed in terms of $\Theta$, the derivatives of the intermolecular potential and $\beta(X)$ [38]. The value of the Lindemann parameter (1), computed at a single melting point $P$, $T_m(P)$, $a(P, T_m)$ that is considered to be known, can be utilized for calculations of the melting curve.

## 3 Results and discussion

To check the accuracy of the Lindemann criterion (1) for strongly anharmonic crystals, we have previously applied it to solid Ar since its melting curve is well known from experiment [33]. For this purpose we have used this parameter calculated at the single experimental melting point $\delta \cong 0.101$. In the range between its triple point $T_t = 83.807$ K and 260 K, the mean-square relative deviation from experimental data makes up about 0.7%.

Using our estimation for the melting temperature of $C_{60}$ fullerite at normal pressure [36], $T_0 = 1500$ K, we calculated the Lindemann parameter at this point: $\delta \cong 0.041$. Then, we solved the equation of state (2) at various fixed pressures up to temperature $T_m(P)$ at which $\sqrt{q^2}/a = 0.041$ [38] and calculated the molar volume at this melting point $V_s = V(P, T_m)$. We have restricted ourselves to a temperature about 4000 K (and a pressure about 15 kbar) since at such temperature the $C_{60}$ molecule is decomposed [29].

Note that in the quasi-harmonic approximation, the mean-square molecular displacements are somewhat different than taking into account the strong anharmonicity. But such an approximation fails at high temperatures. It will suffice to mention [15] that for the temperature of the loss of thermodynamic stability of the two-phase system $C_{60}$ fullerite–vapor, $T_s$, it gives 1054 K, which is much below the estimations for its triple point. Besides, this fullerite was observed at higher temperatures. At the same time Eq. (2) yields $T_s \cong 1916$ K.

The results of our calculations are shown in Fig. 1. The temperature dependence of the melting pressure is described very well by the Simon equation

$$\frac{(P_m(T)/\text{bar})-1}{b} = \left(\frac{T}{T_0}\right)^c. \qquad (6)$$

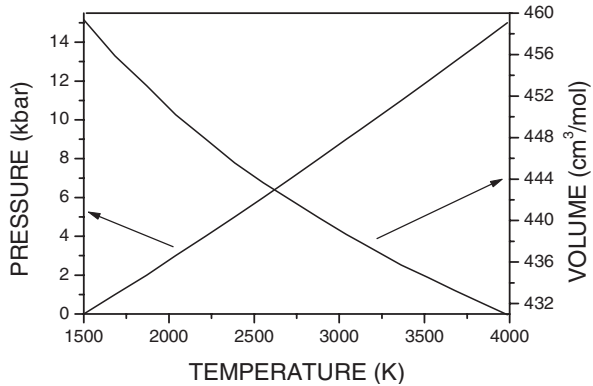

**Fig. 1** The possible melting curve $P_m = P_m(T)$, $V_s = V_s(T)$ of $C_{60}$ fullerite.

Originally, this equation has been proposed for the melting curve of Ar (see e.g. [33]). For $C_{60}$ fullerite we find $T_0 = 1500$ K, $b = 6643.8$, $c = 1.209$. The temperature dependence of the molar volume along the melting curve is approximated by the formula

$$V_s(T) = V_s(T_0) - 29.20 \ln(T/T_0) . \tag{7}$$

It has been demonstrated with Ar [39] that the logarithmic relationship between the molar volume of the solid phase and the temperature along its melting curve is a consequence of the facts that at this curve,

$$\sigma(T, V_s) = S(T, V_s) - S_{id} = S - Nk\left\{\frac{5}{2} + \ln\left[v_s\left(\frac{mkT}{2\pi\hbar^2}\right)^{3/2}\right]\right\} = \text{const} \tag{8}$$

and

$$c_v/\gamma = \text{const} \tag{9}$$

where, in this case, $S$ is the lattice part of the entropy of the crystal, $c_v = C_v^1 - 3R/2$, $C_v^1$ is the lattice part of the isochoric specific heat and $\gamma = (\partial P/\partial T)_v - R/V$. Actually, along the curve calculated, $c_v = \text{const}$ and $\gamma = \text{const}$. Because of this, we can state with assurance that at the melting curve of $C_{60}$ fullerite (6), (7) evaluated using the Lindemann criterion (1), the entropy rule (8) is obeyed as well.

In conclusion, it may be said once more that the liquid phase of fullerites has not hitherto been observed experimentally and we have used the estimation for the melting point of $C_{60}$ fullerite at normal pressure [36] $T_0 = 1500$ K. The experimental value of this temperature in the future may be somewhat different. In such a situation, it will be an easy matter to improve our calculations. It is inconceivable also that the evaluated curve (6), (7) lies on the metastable region of the solid and liquid phases. Nevertheless, the study of the laws of equilibrium between metastable phases of various materials is of significant interest to statistical thermodynamics [40–42] because sometimes two phases that are metastable with respect to a third phase may coexist in equilibrium with one another [40] and it is important to investigate their properties.